\documentclass[aps,pra,showpacs,amsmath,amssymb,amsfonts,superscriptaddress,twocolumn]{revtex4-1}

\usepackage{graphicx}
\usepackage{dcolumn}
\usepackage{bm}
\usepackage{epsf}
\usepackage{amssymb}
\usepackage{color}
\usepackage[colorlinks=true,citecolor=blue,linkcolor=blue,urlcolor=blue]{hyperref}

\newcommand{\bra}[1]{\left\langle #1\right|}
\newcommand{\ket}[1]{\left|#1\right\rangle}

\newcommand{\tr}[1]{\mathrm{tr}\left\{#1\right\}}

\newcommand{\la}{\left\langle}
\newcommand{\ra}{\right\rangle}
\newcommand{\pd}{\partial}
\newcommand{\de}[1]{\delta\left(#1\right)}
\newcommand{\td}{\mathrm{d}}

\newcommand{\id}{\mathbb{I}}
\newcommand{\com}[2]{\left[#1,\,#2\right]}

\newcommand{\co}[1]{\cos{\left(#1\right)}}
\newcommand{\si}[1]{\sin{\left(#1\right)}}

\newcommand{\bla}{bla\\bla\\bla\\bla\\bla}
\newcommand{\PRA}{Phys. Rev. A}

\newcommand{\PRE}{Phys. Rev. E}
\newcommand{\PRL}{Phys. Rev. Lett.}
\newcommand{\PRX}{Phys. Rev. X}

\newcommand{\NJP}{New. J. Phys.}

\newcommand{\mb}[1]{\mbox{\boldmath$#1$}}
\newcommand{\mc}[1]{\mathcal{#1}}
\newcommand{\mbb}[1]{\mathbb{#1}}

\newcommand{\mrm}[1]{\mathrm{#1}}

\begin{document}

\title{Optimal control of a qubit in an optical cavity} 

\author{Sebastian Deffner}
\address{Department of Chemistry and Biochemistry and Institute for Physical Science and Technology, University of Maryland, 
College Park, Maryland 20742, USA}
\email{sebastian.deffner@gmail.com}
\date{\today}

\begin{abstract}
We study quantum information processing by means of optimal control theory. To this end, we analyze the damped Jaynes-Cummings model, and derive optimal control protocols that minimize the heating or energy dispersion rates, and controls that drive the system at the quantum speed limit. Special emphasis is put on analyzing the subtleties of optimal control theory for our system. In particular, it is shown how two fundamentally different approaches to the quantum speed limit can be reconciled by carefully formulating the problem.
\end{abstract}

\pacs{02.30.Yy, 03.67.Ac}

\maketitle

\section{Introduction}

One of the main goals of nanoengineering and quantum optics is the development of nanodevices that reliably process quantum information. A basic requirement for these quantum information processing devices is the ability to universally control the state of a single qubit on timescales much shorter than the coherence time. Promising candidates have been studied experimentally, for instance, in superconducting qubits \cite{li_2013}, quantum-dot charge qubits \cite{cao_2013}, and in cavity QED \cite{cimmarusti_2013}. As in all technological applications the natural question arises how these devices can be operated ``optimally''. 

In this context, an important question in the field of quantum information and quantum (control-)dynamics has recently attracted a lot of attention, namely the \textit{quantum speed limit} \cite{caneva_2009,caneva_2011,hegerfeldt_2013,mukherjee_2013,caneva_2013,def11a,deffner_lutz_2013,campo_2013_qsl,
taddei_2013,barnes_2013,xu_2014,xu_2014a,andersson_2013}. The quantum speed limit time is the minimal time a quantum system needs to evolve between an initial and a final state, and it can be understood as a generalization of the Heisenberg uncertainty relation for time and energy.

A particularly useful set of mathematical tools for approaching this kind of problems is summarized under the headline \textit{optimal control theory}. However, depending on how these tools are used different ``optimal'' results are concluded, which was recently discussed carefully for qubits evolving under unitary dynamics in Ref.~\cite{poggi_2013}. For instance, Caneva \textit{et al.} showed that the Krotov algorithm \cite{caneva_2009} fails to converge if one tries to drive a qubit faster than an independently determined quantum speed limit, while Hegerfeldt used optimal control theory to compute a quantum speed limit that allows even faster evolution \cite{hegerfeldt_2013}. However, quantum  optimal   control theory is not restricted to determining the maximal speed, and has, e.g., been successfully applied to finding driving protocols that maximize squeezing \cite{galve_2009} and entanglement \cite{galve_2009_a} in harmonic oscillators, or efficiently cool molecular vibrations \cite{reich_2013}.

Nevertheless, optimal control theory has remained a mathematical tool box, which is mostly applied in various fields of engineering and applied mathematics, see for instance Refs.~\cite{salamon_2012,boscain_2012,nanduri_2013}, while it is still rather scarcely discussed in the physics' literature and textbooks. However, finding ``optimal'' processes has been an important topic of constant interest in virtually all fields of physics. Only recently, optimal processes in thermodynamic applications have attracted renewed interest \cite{schmiedl_2007,salamon_2009,def10,andresen_2011,sivak_2012,deffner_2013}. Moreover, in quantum computing so called \textit{shortcuts to adiabaticity}  have been in the focus of intense research efforts. These shortcuts are optimal driving protocols that reproduce in a finite time the same outcomes as resulting from an infinitely slow process, see for instance Ref.~\cite{torrontegui_2013,deffner_2014} and references therein.

The purpose of the present paper is two-fold. On the one hand, we will be interested in solving an interesting and important problem, namely how to ``optimally'' control a simple quantum information device. To this end, we will analyze the damped Jaynes-Cummings model by means of optimal control theory, and discuss the optimal finite-time processing of one qubit of information. A similar classical problem was recently analyzed in \cite{zulkowski_2013}.

On the other hand, this paper is also of pedagogical value. We will use the fully analytically solvable example in order to illustrate concepts of optimal control theory, and to ``translate'' between the language typically used in engineering textbooks and vocabulary that is more familiar in quantum thermodynamics. We will illustrate that the formulation of the problem is crucial, as the resulting optimal protocol intimately depends on the question asked. As an important consequence of our study, we will be able to reconcile two fundamentally different approaches to the quantum speed limit.

\paragraph*{Outline}

We aim at a presentation of the results, which is as self-contained as possible. To this end, the paper is organized as follows: we start in Sec.~\ref{sec:opt} with a brief review of elements of optimal control theory, and establish notation. Section~\ref{sec:qubit} is dedicated to a description of the system under study, namely the damped Jaynes-Cummings model. In Sec.~\ref{sec:control} we will derive ``optimal'' control protocols, that minimize the heating rate or minimize the energy dispersion rate. In Sec.~\ref{sec:qsl} we turn to controlling the system at the quantum speed limit. Finally, in Sec.~\ref{sec:conclusion} we conclude the paper with a few remarks.

\section{Elements of Optimal Control Theory\label{sec:opt}}

We start by summarizing the elements of optimal control theory, which we will be using in the following, see also \cite{kirk_2004,dallesandro_2008}. Particular focus will be put on some subtleties that will become important in the later analysis.

Imagine a physical system whose state is fully described by a vector $\mb{x}_t$. The components of $\mb{x}_t$ could be the real, physical microstate, a point in phase space, the state of a qubit, or a collection of macroscopic variables as, for instance, voltage, current, volume, pressure, etc. The evolution of $\mb{x}_t$ for times $0\leq t\leq \tau$ is described by a first order differential equation, the so-called \textit{state equation},
\begin{equation}
\label{q01}
\dot{\mb{x}_t}=\mb{f}\left(\mb{x}_t,\mb{\alpha}_t\right) \quad \mathrm{and}\quad \mb{x}_{t=0}=\mb{x}_0\,,
\end{equation}
where the vector $\mb{\alpha}_t$ is a collection of external control parameters, or simply the control. In a thermodynamic set-up $\mb{\alpha}_t$ can be typically related to a collective degree of freedom of a work reservoir \cite{callen_85,deffner_jarzynski_2013}.

\paragraph*{Accessibility}
A central issue in the set-up of a problem in optimal control theory is the \textit{accessibility}. Note that in mathematical control theory the concept of accessibility is slightly more general than used in the present context. Generally, accessibility refers to controls that are able to drive the state, $\mb{x}_{t}$, to an open set in state space \cite{kirk_2004,dallesandro_2008}. For the sake of simplicity we focus here on the simpler question, namely: given a state equation \eqref{q01} with initial value $\mb{x}_0$, which control protocols $\mb{\alpha}_t$ drive the system to a \textit{specific} state $\mb{x}_{t=\tau}$ during time $\tau$? 

Imagine, for instance, that a qubit is initially prepared in its up-state, and one wants to drive the qubit into its down-state at $t=\tau$. Then only certain parametrizations of an external magnetic field realize this process. In addition, one could imagine that there are further physical constraints, which have to be met by $\mb{\alpha}_t$, as there could be, e.g., limited resources or technical limitations.

This leaves us with a set of \textit{physically allowed} or \textit{admissible} control protocols,
\begin{equation}
\mc{A}=\{\mb{\alpha}_t| \mathrm{admissible\,\, protocols} \}\,.
\end{equation}
However, not all admissible protocols are necessarily practical or even physically meaningful. Thus, we can imagine that some controls fit our purposes better and some worse, and we want to identify the \textit{optimal admissible control} $\mb{\alpha}^*_t\in \mc{A}$.

\paragraph*{Optimal protocols}

In the paradigm of optimal control theory the task is, then, to find the particular $\mb{\alpha}^*_t$ such that a \textit{performance measure}, or \textit{cost functional} is minimized. The cost, $\mc{J}\left[\mb{x}_t,\mb{\alpha_t}\right]$, is usually written as
\begin{equation}
\label{q02}
\mc{J}\left[\mb{x}_t,\mb{\alpha}_t\right]=\int_{0}^\tau \td t\,\mc{L} \left(\mb{x}_t,\dot{\mb{x}}_t,\mb{\alpha}_t\right)\,.
\end{equation}
To find the \textit{optimal control} $\mb{\alpha}^*_t$ we have to minimize the cost functional $\mc{J}\left[\mb{x}_t,\mb{\alpha_t}\right]$ under the condition that $\mb{x}_t$ evolves under the state equation \eqref{q01}. This problem is very similar to problems in classical mechanics \cite{goldstein_1959}, if we identify $\mc{J}\left[\mb{x}_t,\mb{\alpha_t}\right]$ as an action and $\mc{L} \left(\mb{x}_t,\dot{\mb{x}}_t,\mb{\alpha}_t\right) $ as the corresponding Lagrangian of the problem. However, in classical mechanics the action is typically a functional of the state variable only, $\mc{J}\left[\mb{x}_t,\mb{\alpha_t}\right]\equiv\mc{J}\left[\mb{x}_t\right]$, and $\mb{\alpha}_t\equiv\mb{\alpha}_0 $ is a parameter, whereas in the present context we are explicitly asking for an \textit{optimal protocol} $\mb{\alpha}^*_t$.

It is worth emphasizing that the definition of $\mc{J}\left[\mb{x}_t,\mb{\alpha}_t\right]$ is particular to the problem, and depends on the specific situation. We will illustrate this point shortly for a concrete example. However, already here it should be clear that the formulation of the problem anticipates what will be considered optimal. Generally, defining the cost functional is a non-trivial task, and we are considering here only the simplest formulation \cite{kirk_2004}.

\subsection{Hamiltonian formulation -- Pontryangin's optimum principle}

Before we move on to a specific system let us briefly outline how to generally solve the problem. In complete analogy to classical mechanics it is typically more practical to use a ``Hamiltonian'' approach. The control Hamiltonian, $\mc{H}\left(\mb{x}_t,\mb{p}_t,\mb{\alpha}_t\right)$, is obtained by a Legendre transform of the control Lagrangian, $\mc{L} \left(\mb{x}_t,\dot{\mb{x}}_t,\mb{\alpha}_t\right)$, and reads,
\begin{equation}
\label{q03}
\mc{H}\left(\mb{x}_t,\mb{p}_t,\mb{\alpha}_t\right)=\mb{p}_t\cdot \mb{f}\left(\mb{x}_t,\mb{\alpha}_t\right)-\mc{L} \left(\mb{x}_t,\mb{f}\left(\mb{x}_t,\mb{\alpha}_t\right),\mb{\alpha}_t\right)\,,
\end{equation}
where we introduced the canonical momentum $\mb{p}_t$, which is also called the \textit{costate}. Accordingly, the \textit{costate equation} is given by
\begin{equation}
\label{q04}
\dot{\mb{p}}_t=-\frac{\pd}{\pd\mb{x}}\,\mc{H}\left(\mb{x}_t,\mb{p}_t,\mb{\alpha}_t\right)\quad \mathrm{and}\quad \mb{p}_{t=\tau}=0\,.
\end{equation}
It is worth emphasizing again that we are here concentrating on the simplest possible case. Generally the final value of the costate $\mb{p}_{t}$ depends on the problem under study. If, for example, one introduces a terminal cost, i.e., a term in Eq.~\eqref{q02} that penalizes controls that drive the state $\mb{x}_t$ ``only close'' to a specific target state, then $\mb{p}_{t=\tau}$ can be written as a gradient of this terminal cost \cite{kirk_2004}.

Notice that the boundary condition \eqref{q04} determines the \textit{final} value of $ \mb{p}_t $ at $t=\tau$, whereas the state equations \eqref{q01} is an \textit{initial} value problem. This splitting of boundary conditions is a consequence of $ \mc{L} \left(\mb{x}_t,\dot{\mb{x}}_t,\mb{\alpha}_t\right)$ and thus $\mc{H}\left(\mb{x}_t,\mb{p}_t,\mb{\alpha}_t\right)$ depending on a \textit{time-dependent} control $\mb{\alpha}_t $ \cite{kirk_2004}, in contrast to problems in classical mechanics, where $\mb{\alpha}_t\equiv \mb{\alpha}_0$ is only a parameter.

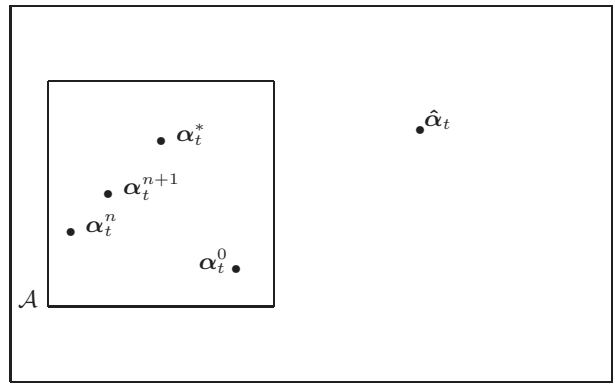
\begin{figure}
\centering
\setlength{\unitlength}{1mm}
\begin{picture}(80,50)
\put(5,10){\line(0,1){30}}
\put(5,10){\line(1,0){30}}
\put(35,10){\line(0,1){30}}
\put(5,40){\line(1,0){30}}
\put(0,0){\line(0,1){50}}
\put(0,0){\line(1,0){80}}
\put(80,0){\line(0,1){50}}
\put(0,50){\line(1,0){80}}
\put(1,10){$\mc{A}$}
\put(25,15){$\mb{\alpha}_t^0$}
\put(30,15){\circle*{1}}
\put(10,20){$\mb{\alpha}_t^n$}
\put(8,20){\circle*{1}}
\put(15,25){$\mb{\alpha}_t^{n+1}$}
\put(13,25){\circle*{1}}
\put(22,32){$\mb{\alpha}_t^*$}
\put(20,32){\circle*{1}}
\put(55,34){$\mb{\hat{\alpha}}_t$}
\put(54.5,33.5){\circle*{1}}
\end{picture}
\caption{\label{fig:admissible} \textbf{Admissible controls as subset of all possible controls:} Illustration of set of admissible controls $\mc{A}$, initial guess, $\mb{\alpha}_t^0$, optimal admissible control $\mb{\hat{\alpha}}_t^* $, optimal control $\mb{\hat{\alpha}}_t $, and two consecutive elements, $\mb{\alpha}_t^n $ and $\mb{\alpha}_t^{n+1} $, of sequence of steepest descent.}
\end{figure}

Pontryagin's principle then states \cite{kirk_2004} that: given $\mb{\alpha}^*_t\in\mc{A} $ is the optimum of all admissible protocol then
\begin{equation}
\label{q05}
\mc{H}\left(\mb{x}^*_t,\mb{p}^*_t,\mb{\alpha}^*_t\right)=\sup_{\mb{\alpha}_t\in \mc{A}}{\mc{H}\left(\mb{x}_t,\mb{p}_t,\mb{\alpha}_t\right)}\,
\end{equation}
where $\sup\{.\}$ denotes the supremum, and $\mb{x}^*_t$ and $\mb{p}^*_t$ are solutions of Hamilton's equation of motion for the optimal control $\mb{\alpha}^*_t$,
\begin{subequations}
\begin{eqnarray}
\label{q06a}
\dot{\mb{x}}^*_t&=\frac{\pd}{\pd\mb{p}}\,\mc{H}\left(\mb{x}^*_t,\mb{p}^*_t,\mb{\alpha}^*_t\right)\quad &\mathrm{and}\quad \mb{x}^*_{t=0}=\mb{x}_0\\
\label{q06b}
\dot{\mb{p}}^*_t&=-\frac{\pd}{\pd\mb{x}}\,\mc{H}\left(\mb{x}^*_t,\mb{p}^*_t,\mb{\alpha}^*_t\right)\quad &\mathrm{and}\quad \mb{p}^*_{t=\tau}=0\,.
\end{eqnarray}
\end{subequations}
If $\mc{H}\left(\mb{x}^*_t,\mb{p}^*_t,\mb{\alpha}^*_t\right)$ is the global maximum for all controls $\mb{\alpha}_t $ then $\mc{H}\left(\mb{x}^*_t,\mb{p}^*_t,\mb{\alpha}^*_t\right)=\mathrm{const.} $, for all times $t$ with $0\leq t\leq\tau$. Generally, however, the global maximum will be attained for a control $\hat{\mb{\alpha}}_t$ that is not admissible, i.e., $\hat{\mb{\alpha}}_t\not\in \mc{A} $, see also Fig.~\ref{fig:admissible} for an illustration. Therefore, we will call the control, $\mc{\mb{\alpha}}^*_t\in\mc{A}$, optimal for which $\mc{H}$ takes a local supremum. Note, for instance, that the supremum could lie on the boundary, $\pd\mc{A} $, of $\mc{A} $. Thus, the control Hamiltonian takes a local \textit{maximum} if and only if $\mc{A} $ is closed, i.e., $ \pd\mc{A}\subset\mc{A}$. We tacitly assumed that the set of admissible controls is not empty, that means that there is at least one admissible control, and therefore a solution exists. 

\subsection{Iterative solution -- Method of steepest descent}

In most situations it is hardly feasible to find analytical expressions for the optimal control $\mb{\alpha}^*_t$. Nevertheless,  the problem can usually be solved iteratively. This means one can find a sequence $\{\mb{\alpha}_t^n\}_{n\in\mbb{N}}$ with $\mb{\alpha}_t^n\in\mc{A} $ which converges towards $\mb{\alpha}_t^*$, that is $\mb{\alpha}_t^n\rightarrow\mb{\alpha}_t^*$ for $n\rightarrow \infty$. To this end, various iterative methods have been developed in numerical approaches, which  typically work for specific sets of problems \cite{kirk_2004,wang_2010,schirmer_2011}. 

One of the conceptually simplest and earliest approaches is the \textit{Method of Steepest Descent} \cite{bryson_1962,bryson_1963,bryson_1964,kirk_2004}. This method is used in virtually all areas of physics, and its convergence properties are well-studied \cite{kirk_2004}. In this approach elements of the sequence, $\{\mb{\alpha}_t^n\}_{n\in\mbb{N}}$, are constructed ``following'' a gradient on a landscape, here along the gradient of the control Hamiltonian with respect to the control. However, for our present purposes we will have to modify the usual algorithm to ensure that all elements of the sequence $\{\mb{\alpha}_t^n\}_{n\in\mbb{N}}$ are actually admissible. 

The \textit{Modified Steepest Descent Algorithm} consists of four steps:
\paragraph*{Step 1:}
Choose  a zeroth order, admissible control, $\mb{\alpha}_t^0\in\mc{A}$, from a sophisticated guess.
\paragraph*{Step 2:} Integrate the state equation \eqref{q01} and obtain a solution $\mb{x}^n_t$, where $n$ is the iteration index of the sequence.
\paragraph*{Step 3:} With $\mb{x}_t=\mb{x}_t^n$ solve the costate equation \eqref{q05}, and evaluate the gradient 
\begin{equation}
\frac{\pd \mc{H}^n}{\pd\mb{\alpha}}=\frac{\pd}{\pd\mb{\alpha}}\,\mc{H}(\mb{x}_t^n,\mb{p}_t^n,\mb{\alpha}_t^n)\,.
\end{equation}
\paragraph*{Step 4:}  Generate a new control as
\begin{equation}
\label{q08}
\mb{\alpha}^{n+1}_t=\mb{\alpha}^n_t+\epsilon \,\frac{\pd \mc{H}^n}{\pd\mb{\alpha}}\,.
\end{equation}
where we introduced the step size $\epsilon$ that has to be determined by some ad hoc strategy and will generally depend on the problem. Generically, there is no guarantee that the such determined control $\mb{\alpha}^{n+1}_t$ is admissible. Therefore, we have to modify the conventional algorithm at this point. As hinted earlier, the modification of the algorithm and hence the optimal control crucially depends on the formulation of the problem. The simplest strategy is to check whether $\mb{\alpha}^{n+1}_t \in \mc{A}$: if the answer is YES then return to \textit{Step 2}, where $\mb{\alpha}^n_t$ is replaced by $\mb{\alpha}^{n+1}_t$; if the answer is NO, return to \textit{Step 2}, where you keep $\mb{\alpha}^n_t$ and choose a new $\epsilon'<\epsilon$ to compute a new $\mb{\alpha'}^{n+1}_t$.

However, this modification is not very systematic and the sequence might easily get ``trapped'', i.e., the algorithm fails to converge. It has been noted that all iterative methods in optimal control theory show this issue of ``getting locally trapped''. One usually has to run the algorithm several times for various combinations of $\mb{\alpha}_t^0\in\mc{A} $ and $\epsilon$  to ensure that the numerical outcome is reliable \cite{kirk_2004}, i.e., the resulting $\mb{\alpha}_t^*$ corresponds to the ``smallest'' local minimum in $\mc{A}$.

Another strategy is to modify the algorithm in a way such that we only have pointwise, but no longer uniform convergence. This can be achieved by introducing a time-dependent stepsize $\epsilon_t$. In simple words this means, that the control sequence $\mb{\alpha}^n_t$ converges towards the optimal control, $\mb{\alpha}^*_t$, with different speeds for different ``points'' $t$ along the protocol. However, also this strategy is not very systematic as it is not ad hoc clear how to find such a time-dependent step size $\epsilon_t$.

Finally, the algorithm is terminated, if the performance measure has sufficiently converged,
\begin{equation}
\label{q09}
\left|\mc{J}\left[\mb{x}^{n+1}_t,\mb{\alpha}_t^{n+1}\right]-\mc{J}\left[\mb{x}^n_t,\mb{\alpha}_t^n\right]\right|\leq \delta\,,
\end{equation}
where $\delta$ is a preselected positive constant.

In the following we will consider a specific problem namely the optimal processing of one qubit of information. We will see that in this case the phrase ``admissible protocol'' can be expressed as a simple mathematical condition, which in turn will allow us to \textit{systematically} modify  Eq.~\eqref{q08} such that $\mb{\alpha}^{n+1}_t\in \mc{A}$ for all $\epsilon\in\mbb{R}$. Before we discuss this point in more detail let us therefore formulate our specific problem, first.

\section{The system -- Damped Jaynes-Cummings model\label{sec:qubit}}

\begin{figure}
\centering
\includegraphics[width=.48\textwidth]{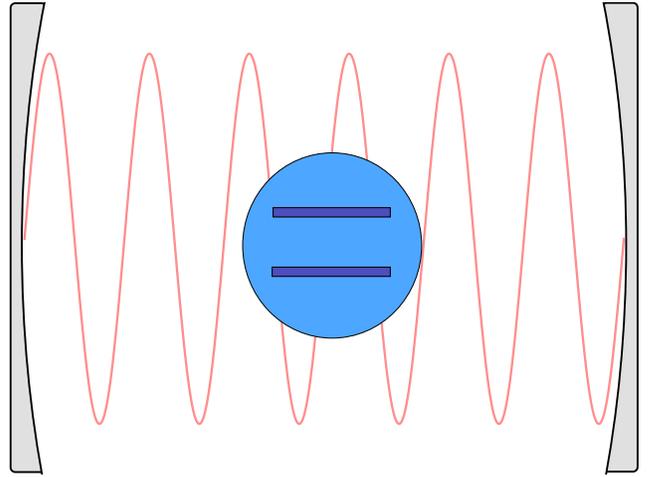}
\caption{\label{fig:cavity} \textbf{(color online) Illustration of the system under consideration:}  A single qubit in an optical cavity interacting with a LASER field.}
\end{figure}

For the remainder of the paper we shall be interested in the optimal processing of one qubit of information. To this end, we analyze a two-level atom dissipatively coupled to a leaky, optical cavity as illustrated in Fig.~\ref{fig:cavity}.

The mathematical description of this system is commonly known as damped Jaynes-Cummings model \cite{breuer_07}, and it will prove useful for our present purposes that the model is fully analytically solvable. This will allow us to illustrate the concepts of optimal control theory and information processing in relatively simple terms and mathematical expressions. 

For the sake of completeness, we summarize the derivation of the dynamics. Consider a \textit{self-contained} universe \cite{deffner_jarzynski_2013}, whose total Hamiltonian reads,
\begin{equation}
\label{q10}
H_\mrm{tot}=H_\mrm{qubit}\otimes\id_\mrm{cavity}+\id_\mrm{qubit}\otimes H_\mrm{cavity}+H_\gamma\,.
\end{equation}
By $H_\mrm{qubit}$ we denote the reduced Hamiltonian of the qubit, $H_\mrm{cavity}$ describes the cavity, and $H_\gamma$ is an interaction term. With the Pauli operators, $\sigma_\pm=(\sigma_x\pm i \sigma_y)/2$, and $\hbar\omega_0$ being the energy difference between ground and excited state, we have for the qubit \cite{breuer_07}
\begin{subequations}
\begin{equation}
\label{q11a}
H_\mrm{qubit}=\hbar \omega_0\,\sigma_+\sigma_-\,.
\end{equation}
The cavity Hamiltonian reads
\begin{equation}
\label{q11b}
H_\mrm{cavity}=\sum_k \hbar \omega_k\,b^\dagger_k b_k\,,
\end{equation}
where $k$ labels the field modes, $\omega_k$ are the cavity frequencies, and $b^\dagger_k, b_k $ are creation and annihilation operators, respectively. Finally the interaction is written as
\begin{equation}
\label{q11c}
H_\mrm{\gamma}=\sigma_+\otimes B+\sigma_-\otimes B^\dagger\quad\mathrm{with}\quad B=\sum_k \gamma_k b_k\,.
\end{equation}
\end{subequations}
By $\gamma_k$ we denoted the coupling constants. If we further assume that the cavity is initially prepared in a vacuum state, then the exact master equation for the reduced density operator of the qubit only, $\rho_t$, can be written as \cite{breuer_07,garraway_1997}
\begin{eqnarray}
\label{q12}
\dot{\rho}_t&=&-\frac{i}{\hbar} \com{H_\mrm{qubit}}{\rho_t}-\frac{i}{2\hbar} \com{\lambda_t\, \sigma_+\,\sigma_-}{\rho_t} \nonumber\\
&+&\gamma_t\left(\sigma_-\rho_t\sigma_- -\frac{1}{2}\sigma_+\sigma_-\rho_t-\frac{1}{2}\rho_t\sigma_+\sigma_-\right)\,.
\end{eqnarray}
The time-dependent decay rate, $\gamma_t$, and the time-dependent Lamb shift, $\lambda_t$, are fully determined by the spectral density, $J(\omega)$, of the cavity mode. We have
\begin{equation}
\label{q13}
\lambda_t=-2\,\mrm{Im}\left\{\frac{\dot{c}_t}{c_t}\right\} \quad \mathrm{and} \quad \gamma_t=-2\,\mrm{Re}\left\{\frac{\dot{c}_t}{c_t}\right\}\,
\end{equation}
where $c_t$ is a solution of
\begin{equation}
\label{q14}
\dot{c}_t=-\int_0^t\td s\int \td \omega\,J(\omega)\,e^{i\hbar\left(\omega-\omega_0\right)(t-s)}\,c_s\,.
\end{equation}
This model has been extensively studied, since it is exact and completely analytically solvable \cite{breuer_07,garraway_1997}. Moreover, it is of thermodynamic relevance as it allows the study of non-Markovian quantum dynamics \cite{breuer_2009,laine_2010,xu_2010,hou_2011,fonseca_2012} and it has recently been realized in a solid-state cavity QED \cite{madsen_2011}.

In the paradigm of optimal control theory $\gamma_t$ and $\lambda_t$ can be interpreted as two control parameters and we have $\mb{\alpha}_t=(\gamma_t,\,\lambda_t)$. Physically the control protocol can be realized by appropriately choosing real and imaginary parts of the Fourier transform of the spectral density of the cavity mode.

\section{Optimal quantum information processing \label{sec:control}}

We now continue to find controls that optimally process one qubit of information, i.e., analyze the damped Janyes-Cummings model by means of optimal control theory. First we have to formulate the problem by deriving the state equation and identifying the set of admissible controls. In the second part of this section we then define various physically relevant performance measures for which we obtain the optimal controls. 

\subsection{Solution of the model -- The state equation}

A convenient way to represent the density operator of a qubit is the Bloch representation,
\begin{equation}
\label{q15}
\rho_t=\frac{1}{2}\left(\id_2+x_t\,\sigma_x+y_t\,\sigma_y+z_t\,\sigma_z\right)\,.
\end{equation}
In this formulation $\rho_t$ is described by a three dimensional vector, $(x_t,y_t,z_t)$, which lives in the Bloch sphere described in Cartesian coordinates by  $\sqrt{x_t^2+y_t^2+z_t^2}\leq 1$. Pure states lie on the surface of the Bloch sphere, whereas the fully mixed states is at the origin.

In this representation Eq.~\eqref{q12} can be written as a system of three linearly coupled differential equations. We have for $x_t$
\begin{subequations}
\label{q16}
\begin{equation}
\label{q16a}
\dot{x}_t=-\frac{\gamma_t}{2}\,x_t-\frac{\lambda_t}{2}\, y_t-\omega_0\, y_t
\end{equation}
while $y_t$ evolves according to
\begin{equation}
\label{q16b}
\dot{y}_t=\frac{\lambda_t}{2}\,x_t+\omega_0\, x_t-\frac{\gamma_t}{2}\, y_t\,.
\end{equation}
The equation of motion for $z_t$ decouples of $x_t$ and $y_t$, and we obtain
\begin{equation}
\label{q16c}
\dot{z}_t=-\gamma_t\, z_t-\gamma_t\,.
\end{equation}
\end{subequations}
The latter three Eqs.~\eqref{q16a}-\eqref{q16c} are readily identified as state equation \eqref{q01}, where the initial value will be set shortly.

With $\Gamma_t=\int_0^t\td s\, \gamma_s$ and $\Lambda_t=\int_0^t \td s\, \lambda_s$ the solution of the state equation \eqref{q16} can be written as
\begin{equation}
\label{q17}
x_t=e^{-\Gamma_t/2}\left(x_0\, \co{\omega_0 t+\Lambda_t/2}-y_0\, \si{\omega_0 t+\Lambda_t/2}\right)\,,
\end{equation}
and 
\begin{equation}
\label{q18}
y_t=e^{-\Gamma_t/2}\left(x_0\, \co{\omega_0 t+\Lambda_t/2}+y_0\,\co{\omega_0 t+\Lambda_t/2}\right)\,.
\end{equation}
Finally, the evolution of the $z_t$-component becomes 
\begin{equation}
\label{q19}
z_t=z_0\,e^{-\Gamma_t}+\left(e^{-\Gamma_t}-1\right)\,,
\end{equation}
which will allow us to represent the set of admissible controls, $\mc{A}$, in a particularly simple way.

\subsection{Accessibility -- Admissible controls}

For the sake of simplicity let us restrict ourselves to specific initial and final states. We assume that initially the qubit is prepared in its up-state, and hence we have 
\begin{equation}
\label{q20}
x_0=0,\quad y_0=0, \quad \mrm{and}\quad z_0=1\,.
\end{equation}
Now we are interested in \textit{writing} one qubit of information and the final state is determined by
\begin{equation}
\label{q21}
x_\tau=0,\quad y_\tau=0, \quad \mrm{and}\quad z_\tau=0\,.
\end{equation}
Note that we are using the terms \textit{writing} and \textit{erasing} of information in its thermodynamic sense. One easily convinces oneself that the von Neumann entropy of the initial state is
\begin{equation}
\label{q22}
\mc{S}_0=-\tr{\rho_0\ln\rho_0}=0
\end{equation}
and we have for the final state
\begin{equation}
\label{q23}
\mc{S}_\tau=-\tr{\rho_\tau\ln\rho_\tau}=\ln 2\,.
\end{equation}
Therefore, the process under consideration increases the von Neumann entropy by $\Delta\mc{S}=\ln 2$, which is interpreted as writing one qubit of information \cite{sha48,ben-naim_08}.

For our specific choice of initial \eqref{q20} and final \eqref{q21} state  the situation greatly simplifies. The state equation effectively reduces to a simple one-dimensional differential equation \eqref{q16c}, and we have only one control, namely the time-dependent decay rate, $\gamma_t$. By further employing the solution for $z_t$ \eqref{q19} and setting $z_\tau=1$, the set of admissible controls can be written as
\begin{equation}
\label{q24}
\mc{A}=\left\{\gamma_t\bigg|\,\int_0^\tau\td t\,\gamma_t=\ln{2}\right\}\,.
\end{equation}
To summarize, for the following analysis we have to consider a single state $z_t$, which evolves according to \eqref{q16c}, with boundary conditions $z_0=1$ and $z_\tau=0$. Further, admissible controls are determined by Eq.~\eqref{q24}, and we can now proceed by \textit{defining} what we will call \textit{optimal}.

\subsection{Minimal heating rate}

As a first example let us imagine that we want to write one qubit of information, while the power exchange with the environment is minimal. This corresponds to the typical situation in experiments, where for various reasons the heating rate, quantified by the power input, shall be kept small. For a qubit interacting with a leaky single-mode cavity \eqref{q12} we have, 
\begin{equation}
\label{q25}
\dot{Q}_t=\tr{H_\mrm{qubit}\,\dot{\rho}_t}=\frac{\hbar\omega_0}{2}\,\dot{z}_t\,,
\end{equation}
where we denoted the heating rate as $\dot{Q}_t$. Generally $\dot{Q}_t$ can have either sign, depending on whether energy is pumped into the qubit or taken out. 

Therefore, we \textit{define} the performance measure to read
\begin{equation}
\label{q26}
\mc{J}_Q[z_t,\,\gamma_t]=\int_0^\tau \td t\,\dot{z}_t^2=\int_0^\tau \td t\, \gamma^2_t\,(z_t+1)^2\,,
\end{equation}
where the square is a convenient choice to ensure that $\mc{J}_Q[z_t,\,\gamma_t]$ is non-negative. Then the control Hamiltonian \eqref{q03} can be written as
\begin{equation}
\label{q27}
\mc{H}_Q(z_t,p_t,\gamma_t)=-\gamma^2_t\,(z_t+1)^2-\gamma_t\,p_t\,(z_t+1)\,,
\end{equation}
and the costate equation becomes
\begin{equation}
\label{q28}
\dot{p}_t=\gamma_t\,p_t+2\,\gamma_t^2\,(z_t+1)\quad \mathrm{and}\quad p_\tau=0\,.
\end{equation}
As the state equation \eqref{q16c} the costate equation \eqref{q28} can be solved analytically, and we have
\begin{equation}
\label{q29}
p_t=-4\,e^{\Gamma_t}\,\int_t^\tau\td s\,\dot{\Gamma}^2_s\,e^{-2\Gamma_s}\,.
\end{equation}
Substituting the solutions \eqref{q19} and \eqref{q29} in the control Hamiltonian \eqref{q27} leaves us with an integro-differential equation for $\gamma_t$. The optimal control $\hat{\gamma}_t$ is determined by finding the particular control(s) for which $ \mc{H}_Q(\hat{\gamma}_t)$ is constant. As argued earlier, $\hat{\gamma}_t$ will generically not be an admissible control, cf. Fig.~\ref{fig:admissible}. Therefore, we continue our analysis with constructing a sequence of admissible controls, $\gamma_t^n\rightarrow \gamma^*_t$, to find the \textit{optimal admissible control}, $\gamma^*_t$.

\paragraph*{Optimal control sequence}

Our aim is to make full use of the Modified Algorithm of Steepest Descent. To this end, let us consider the gradient,
\begin{equation}
\label{q30}
\frac{\pd \mc{H}_Q}{\pd\gamma}=-2\,\gamma_t\,(z_t+1)^2-p_t\,(z_t+1)\,.
\end{equation}
As noted above, if we construct a sequence naively as,
\begin{equation}
\label{q31}
\gamma^{n+1}_t=\gamma_t^n+\epsilon \,\frac{\pd \mc{H}^n}{\pd\gamma}\,,
\end{equation}
then typically $\gamma^{n+1}_t $ will not be an admissible protocol \eqref{q24}, that is here
\begin{equation}
\label{q32}
\int_0^\tau\td t\,\gamma^{n+1}_t\neq\ln 2\,.
\end{equation}
This means that we have to modify the usual algorithm in a way that integral remains invariant, 
\begin{equation}
\label{q34}
\int_0^\tau\td t\,\gamma^{n}_t=\int_0^\tau\td t\,\gamma^{m}_t\quad\mathrm{for\hspace{.25em} all} \quad n,m\in\mbb{N}\,.
\end{equation}
Therefore, a modified sequence can be constructed as,
\begin{equation}
\label{q33}
\gamma^{n+1}_t=\gamma_t^n+\epsilon \,\left(\frac{\pd \mc{H}^n}{\pd\gamma}-\int_0^\tau\td t\,\frac{\pd \mc{H}^n}{\pd\gamma}\right)\,,
\end{equation}
for which, with $\gamma_t^0\in\mc{A}$, all controls of the sequence, $\gamma_t^n $, are admissible. One easily convinces oneself that the modified sequence still converges uniformly as $\int_0^\tau\td t\,\pd \mc{H}^n/\pd\gamma$ is simply a numerical constant. 

\begin{figure}
\centering
\includegraphics[width=.48\textwidth]{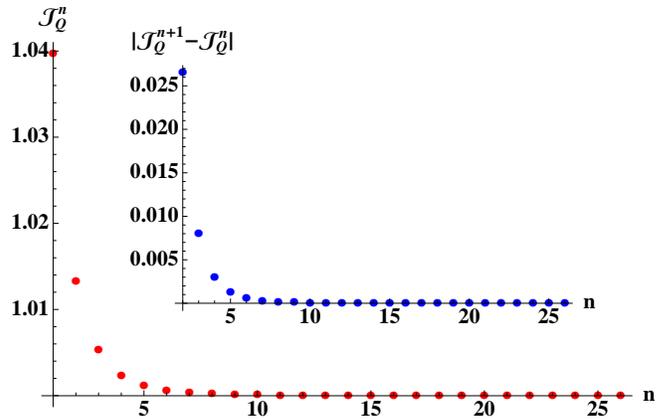}
\caption{\label{fig:J_heat}\textbf{(color online) Cost functional for minimal heating rate:} Cost functional \eqref{q26} for the modified algorithm \eqref{q33} for Method of Steepest Descent with constant control \eqref{q35} as initial guess, stepsize $\epsilon=0.1$, and termination parameter $\delta=10^{-5}$}
\end{figure}

\begin{figure}
\centering
\includegraphics[width=.48\textwidth]{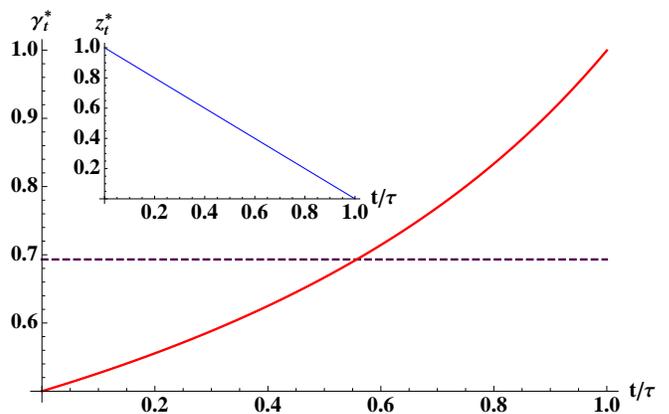}
\caption{\label{fig:z_heat}\textbf{(color online) Optimal admissible control for minimal heating rate:} Optimal admissible control $\gamma_t^*$ (red, solid line) together with initial, sophisticated guess $\gamma_t^*$ \eqref{q35} (purple, dashed line); optimal evolution of state $z^*_t$ as an inset.}
\end{figure}

The simplest admissible control is described by a constant protocol, which we choose as our initial sophisticated guess,
\begin{equation}
\label{q35}
\gamma_t^0=\ln 2/\tau\,.
\end{equation}
In Fig.~\ref{fig:J_heat} we plot the sequence of performance measures with a stepsize $\epsilon=0.1$. We observe that the algorithm converges within the first 25 iterations \footnote{The algorithm was also run for various other initial, admissible controls, and the same convergence behavior was observed. Moreover, it was numerically checked that the resulting optimal admissible control is independent of the initial guess.}. Note that ``convergence'' is quantified by Eq.~\eqref{q09}, that means a sequence is considered to ``have converged'' if the inequality \eqref{q09} is fulfilled. For the specific example in Fig.~\ref{fig:J_heat} we chose $\delta=10^{-5}$. 
Figure~\ref{fig:z_heat} shows the optimal admissible control $\gamma_t^*$ together with the optimal trajectory $z_t^*$. It turns out that in the optimal case $z_t^*$ is a linearly decreasing function, for which the heating rate \eqref{q25} is negative and constant.

\subsection{Minimal energy dispersion}

\begin{figure}
\centering
\includegraphics[width=.48\textwidth]{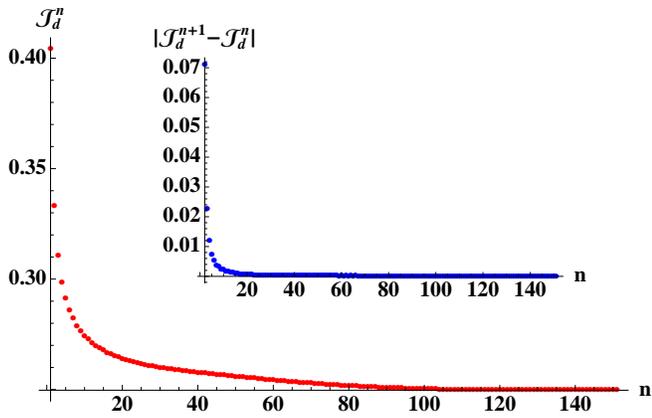}
\caption{\label{fig:J_dis}\textbf{(color online) Cost functional for minimal dispersion rate:} Cost functional \eqref{q37} for the modified algorithm \eqref{q33} for Method of Steepest Descent with constant control \eqref{q35} as initial guess, stepsize $\epsilon=0.25$, and termination parameter $\delta=10^{-5}$.}
\end{figure}

\begin{figure}
\centering
\includegraphics[width=.48\textwidth]{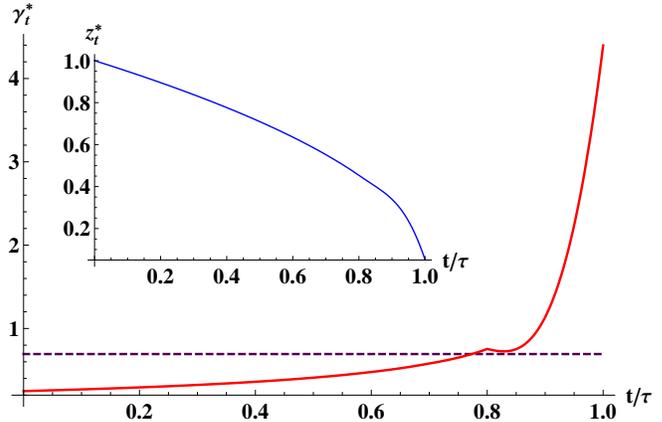}
\caption{\label{fig:z_dis}\textbf{(color online) Optimal admissible control for minimal dispersion rate:} Optimal admissible control $\gamma_t^*$ (red, solid line) together with initial, sophisticated guess $\gamma_t^*$ \eqref{q35} (purple, dashed line); optimal evolution of state $z^*_t$ as an inset.}
\end{figure}

As a second example let us consider processes were we want to minimize the rate with which the internal energy of the qubit disperses. Such ``optimal'' protocols might be important in situations where one has to worry about decoherence due to some additional coupling to the environment. To this end, consider the variance of the Hamiltonian
\begin{equation}
\label{q36}
\la H_0^2\ra-\la H_0\ra^2=\frac{\hbar^2\omega_0^2}{4}\,\left(1-z_t^2\right)\,,
\end{equation}
from which we compute the rate of dispersion as, $d_t=\td/\td t\,(\la H_0^2\ra-\la H_0\ra^2)$. In this case the performance measure \eqref{q02} can be defined as
\begin{equation}
\label{q37}
J_d[z_t,\,\gamma_t]=\int_0^\tau \td t\,d_t^2=\int_0^\tau \td t\, \gamma_t^2\,(1+z_t)^2\, z^2_t\,.
\end{equation}
Accordingly, the control Hamiltonian becomes
\begin{equation}
\label{q38}
\mc{H}_d(z_t,p_t,\gamma_t)=-\gamma_t^2\,(1+z_t)^2\, z^2_t-\gamma_t\,p_t\,(z_t+1)\,,
\end{equation}
which yields the costate equation $\dot{p}_t=-\pd \mc{H}_d/\pd z$ and the gradient $\pd \mc{H}_d/\pd\gamma$ necessary to construct the optimal control sequence $\gamma_t^n$. In Fig.~\ref{fig:J_dis} we plot the resulting sequence of performance measures, where we observe that the convergence is much slower than in the case of the minimal heating rate, cf. Fig.~\ref{fig:J_heat}.
Nevertheless, the sequence converges satisfactorily and the resulting admissible optimal control is shown in Fig.~\ref{fig:z_dis} \footnote{The algorithm was again run for various other initial controls and  it was numerically checked whether the resulting optimal admissible control is independent of the initial guess and reliable.}. We observe that the optimal control with minimal dispersion rate is significantly different from the control that minimizes the heating rate. The small knick around $t\simeq 0.8 \tau$ is most likely a numerical artefact, which probably could be ``ironed out'' by letting the algorithm run for a longer period. However, it seems that this artefact is a generic peculiarity of the problem, as it appears for various initial controls $\gamma_t^0$. 

Comparing Figs.~\ref{fig:z_heat} and \ref{fig:z_dis} illustrates our earlier point, namely that the resulting optimal control crucially depends on the set-up and formulation of the problem.

\section{Driving at the quantum speed limit\label{sec:qsl}}

In the previous sections we introduced elements of optimal control theory and a conceptually simple model system for quantum information processing. In addition, we illustrated concepts and methods by deriving the optimal control protocols, which either minimize the heating rate or the energy dispersion rate. Equipped with these methods we now continue to analyze the problem of processing information at the quantum speed limit.

For uncontrolled, time-independent systems, the quantum speed limit time determines the maximum rate of evolution, and is a bound combining the results of Mandelstam-Tamm (MT) \cite{man45} and Margolus-Levitin (ML) \cite{mar98}: it is given for isolated, time-independent systems by $\tau_\mrm{QSL}= \mbox{max}\{\pi\hbar/(2\Delta E),\pi \hbar/(2E)\}$, where $\Delta E$ is the variance of the energy of the initial state and $E$ its mean energy with respect to the ground state. Generalizations of the MT and ML findings to driven and open systems have been recently proposed in Refs.~\cite{def11a,taddei_2013,campo_2013_qsl,deffner_lutz_2013}. The approach in these papers has been called geometric, as the derivation relies on an estimation of the geometric speed.

\paragraph*{Geometric approach} 
The question one asks is the following: given a particular external control, how fast can a quantum system follow? The answer is given by the maximal speed of quantum evolution. To this end, consider the evolution from an initially pure state $\rho_0=\ket{\psi_0}\bra{\psi_0}$  to a final state $\rho_\tau$.  Under  nonunitary dynamics,  the final state $\rho_\tau$ will be generally  mixed. The geometric approach is then  based on the dynamical properties of the Bures angle $\ell(\rho_0,\rho_\tau)$  between initial and final states of the quantum system \cite{bur69,joz94}, 
\begin{equation}
\label{q39}
\ell(\rho_0,\rho_\tau)=\arccos\left(\sqrt{\bra{\psi_0}\rho_\tau\ket{\psi_0}}\right).
\end{equation}
The Bures angle is a generalization to mixed states of the angle in Hilbert space between two state vectors \cite{nielsen_00}.

The maximal speed of quantum evolution is then determined by \cite{deffner_lutz_2013}
\begin{equation}
\label{q40}
2\co{\ell}\si{\ell}\,\dot{\ell}\leq ||\dot{\rho}_t||_\mrm{op}\,,
\end{equation}
where $||\cdot||_\mrm{op}$ denotes the operator norm, i.e., the largest singular value.

\paragraph*{Minimal time approach}

A fundamentally different question was addressed in Ref.~\cite{hegerfeldt_2013}, namely: what is the minimal time a qubit needs to evolve from \textit{particular} initial state to a \textit{particular} final state? Moreover, it was shown for a qubit evolving under unitary dynamics \cite{hegerfeldt_2013} that this problem can be solved by means of optimal control theory. Since the geometric approach and the minimal time approach yield the same quantum speed limit for time-independent systems, it was not ad hoc clear which approach is more physically relevant for driven systems \cite{poggi_2013}.

\paragraph*{Importance of the formulation}
We have already seen earlier that the formulation of the problem anticipates what will be considered optimal. In particular, the choice of admissible controls is crucial, and a more careful analysis  of the formulation and set-up to derive quantum speed limits is in order. Therefore, we continue our analysis by deriving the optimal controls resulting from the minimal time approach and from the geometric approach for our model system introduced above. It will turn out, that by carefully formulating the problem, both approaches, minimal time and geometric, can be reconciled.

\subsection{Minimal time approach}

In the minimal time approach one is interested in minimizing the process time $\tau$, during which the qubit evolves. Therefore, the performance measure \eqref{q02} is simply given by
\begin{equation}
\label{q41}
\mc{J}_\tau[z_t,\,\gamma_t]=\tau\,.
\end{equation}
In this case $\mc{J}_\tau[z_t,\,\gamma_t]$ is no longer a functional of the control $\gamma_t$ and the state $z_t$, but reduces to a parameter. Accordingly, the control Hamiltonian reads
\begin{equation}
\label{q42}
\mc{H}_\tau(z_t,p_t,\gamma_t)=-\tau -\gamma_t\,p_t\,(z_t+1)\,,
\end{equation}
and, hence, we obtain for the costate equation
\begin{equation}
\label{q43}
\dot{p}_t=\gamma_t\,p_t\quad\mathrm{and}\quad p_\tau=0\,.
\end{equation}
The latter differential equation can be readily solved analytically and the solution is written as
\begin{equation}
\label{q44}
p_t=p_0\,e^{\Gamma_t}\quad\Rightarrow\quad p_t\equiv 0\,.
\end{equation}
It is easy to see that for all real $\gamma_t$ the boundary condition can only be fulfilled for $p_0=0$, and therefore we have $p_t\equiv 0$. Thus, the control Hamiltonian for this problem \eqref{q42} greatly simplifies. Since $\tau\geq0$ and $\mc{H}_\tau(z^*_t,p^*_t,\gamma^*_t)$ is maximal for the optimal control \eqref{q05}, we conclude
\begin{equation}
\label{q43a}
\mc{H}_\tau(z^*_t,p^*_t,\gamma^*_t)=-\tau^* \quad\Rightarrow\quad \tau^*=0\,.
\end{equation}
One easily convinces oneself that the optimal admissible control then has to read
\begin{equation}
\label{q44a}
\gamma_t^*=\ln{2}\,\de{t}\,.
\end{equation}
From these results one is tempted to conclude that a qubit coupled to a leaky cavity can evolve arbitrarily fast, and there is no fundamental bound on the minimal evolution time. However, it is not immediately clear if this result is only of mathematical nature, or if it is also physically relevant, see also the discussion in \cite{poggi_2013}.

In a previous section we considered the heating rate, which quantifies the power input into the qubit. For the latter optimal control \eqref{q44} the state trajectory \eqref{q19} is described by $z^*_t=\Theta(-t)$, and hence we obtain
\begin{equation}
\label{q45}
\dot{Q}^*_t=\frac{\hbar\omega_0}{2}\ln{2}\,\left(\Theta(-t)+1\right)\,\de{t}\,.
\end{equation}
One easily convinces oneself, that the latter expression is the \textit{maximal} heating rate, i.e., the maximal power input possible. 

Equation~\eqref{q45} clearly shows, that for our system infinitely fast evolution is possible if at the same time infinitely much power is pumped into the qubit. However, this process will generally not be relevant for any experimental situation, as an infinite power impulse very likely will ``kick'' the atom out of the optical cavity. 

\subsection{Geometric approach -- Maximal quantum speed}

Now let us turn to the geometric approach. To this end, we evaluate the Bures angle \eqref{q39}, which simply reads
\begin{equation}
\label{q46}
\ell(\rho_0,\rho_t)=\arccos\sqrt{\frac{1}{2}\,\left(1+z_t\right)}\,.
\end{equation}
First, we have to ensure that the inequality in Eq.~\eqref{q40} is a tight bound. The left side of inequality \eqref{q40} becomes
\begin{equation}
\label{q47}
2\co{\ell}\si{\ell}\,\dot{\ell}=-\dot{z}_t/2\,,
\end{equation}
and we have for the right side
\begin{equation}
\label{q48}
||\dot{\rho}_t||_\mrm{op}=\left|\dot{z}_t\right|/2\,.
\end{equation}
In order to formulate the performance measure $\mc{J}_\mrm{QSL}[z_t,\,\gamma_t]$ \eqref{q02} we will, therefore, need a term $\left(|\dot{z}_t|+\dot{z}_t\right)^2$, that minimizes the difference between left and right side and ``makes'' the inequality \eqref{q40} as close as possible to an equality. 

Second, we are interested in such processes, whose evolution speed is maximal, i.e., $|\dot{z}|$ is maximal. Therefore, a performance measure $\mc{J}_\mrm{QSL}[z_t,\,\gamma_t]$ can be defined to read
\begin{eqnarray}
\label{q49}
\mc{J}_\mrm{QSL}[z_t,\,\gamma_t]&=\int_0^\tau\td t\,\left[\left(|\dot{z}_t|+\dot{z}_t\right)^2-|\dot{z}|^2\right]\nonumber \\
&=-\int_0^\tau\td \,\gamma_t^2\left(z_t+1\right)^2\,,
\end{eqnarray}
which we immediately recognize as minus the performance measure that minimizes the heating rate \eqref{q26}, $\mc{J}_\mrm{QSL}[z_t,\,\gamma_t]=-\mc{J}_Q[z_t,\,\gamma_t] $. The optimal admissible protocol that maximizes the quantum evolution speed is identical to the control that maximizes the heating rate. Such an optimal control, however, is just the optimal protocol that we derived within the minimal time approach, namely the delta-peak control \eqref{q44}. 

In conclusion, we explained that the minimal time approach and geometric approach ask fundamentally different questions. However, we also showed that if the problem is formulated carefully by means of optimal control theory, the same results for the quantum speed limit time can be obtained. We found that there is no fundamental bound on the speed with which one qubit of information can be written by a leaky, optical cavity, if we allow for an infinite power input into the system. Practically, however, one is rather interested in ``optimal'' controls, that are more experimentally relevant, as for instance, the fastest evolution under a bounded heating rate. To solve these problems one first has to carefully define the admissible controls, and find a cost functional the reflects the full physical situation. We expect that the actual quantum speed limit is then governed by, for instance, the maximal control power.

Naively computing quantum speed limits by means of optimal control theory can yield unphysical results. Therefore, special attention has to be paid to a careful definition of the set of admissible controls and the performance measure. The outcome of optimal control theory is only as good, i.e., physical as the formulation of the problem.

\section{Concluding remarks\label{sec:conclusion}}

In this paper we have shown how to find control protocol that optimally process one qubit of information. To this end, we have presented some elements of optimal control theory. For a specific system, namely the damped Jaynes-Cummings model, we then have developed a Modified Method of Steepest Descent, which ensures that all elements of a control sequence are actually admissible controls. With this novel algorithm we have numerically determined the optimal controls that minimize the power input and the dispersion rates.

Special emphasis has been put on illustrating that the outcome of an analysis by means of optimal control theory crucially depends on the formulation of the problem. In doing so, we have been able to reconcile two fundamentally different approaches to the quantum speed limit, which yield the same result if the problem is formulated carefully.

Last but not least this paper is of pedagogical value. The presentation of the analysis is mostly self-contained and we hope that our results will spur interactions between different fields. In particular, we believe that this paper could make optimal control theory more accessible and known among physicists, and introduce engineers and applied mathematicians to problems and questions in quantum thermodynamics.

\acknowledgements{It is a pleasure to thank Marcus Bonan\c{c}a and Andrew Smith for stimulating discussions. We acknowledge financial support from the National Science Foundation (USA) under grant DMR-1206971.}


%

\end{document}